\documentclass[aip,jap,amssymb,amsmath]{revtex4-1}%
\usepackage{graphicx}
\usepackage[colorlinks=true,linkcolor=blue]{hyperref}%

\begin{document}

\title{Ferroelectric nanodomains in epitaxial PbTiO$_3$ films grown on SmScO$_3$ and TbScO$_3$ substrates}

\author{F. Borodavka}
\author{I. Gregora}
\affiliation{Institute of Physics, Academy of Sciences of the Czech Republic\\%
Na Slovance 2, 182 21 Prague 8, Czech Republic\\}
\author{A. Bartasyte}
\affiliation{Institute Jean Lamour, CNRS-UMR 7198, Universit\'e de Lorraine, Parc de Saurupt, CS14234, F-54042 Nancy, France}
\author{S. Margueron}
\affiliation{Laboratoire Mat\'eriaux Optiques, Photoniques et Syst\`{e}mes, EA 4423, Universit\'e de Lorraine and Sup\'elec, 2 rue E. Belin, 57070 Metz, France}
\author{V. Plausinaitiene}
\author{A. Abrutis}
\affiliation{University of Vilnius, Department of General and Inorganic Chemistry, Naugarduko 24, LT-03225, Vilnius, Lithuania}
\author{J. Hlinka}
\email{hlinka@fzu.cz}
\affiliation{Institute of Physics, Academy of Sciences of the Czech Republic\\%
Na Slovance 2, 182 21 Prague 8, Czech Republic\\}

\date{\today}

\begin{abstract}
Domain structures of 320\,nm thin epitaxial films of ferroelectric
PbTiO$_3$ grown by MOCVD technique in identical conditions on
SmScO$_3$ and TbScO$_3$ perovskite substrates have been
investigated by Raman spectroscopy and piezoresponse force
microscopy techniques. Phonon frequency shifts and typical domain structure motifs are discussed. 
The results reveal strikingly different domain structure architecture: domain structures of the PbTiO$_3$
film grown on SmScO$_3$ have dominantly $a$-domain orientation
while strongly preferential $c$-domain orientation was found in
the PbTiO$_3$ film grown on the TbScO$_3$ substrate. 
Differences between the two cases are traced back to the film-substrate 
lattice mismatch at the deposition temperature.
\end{abstract}

\pacs{77.80.-e,77.80.Dj,68.55.-a,68.37.Ps}

\maketitle

\section{Introduction}
Morphology of domain structure in thin films of tetragonal
ferroelectric perovskite oxides has recently attracted a
considerable amount of interest.\cite{Book2,Ganpule02,
Roytburd01,Alpay98III,Koukhar01,Pertsev98,streiff02, Schilling06}
 Various special patterns were seen, such as ferroelectric flux closure
 and quadrupolar vortex arrangements.\cite{Fu03,Scott05,Prosandeev06,Gruverman08,Rodriguez09,McGillyNano10,Ivry10,McGillyJMS09,SchillingPRB11,SchillingNano09,McQuaid11}
Most peculiar domain arrangements were found in theoretical model
simulations conducted for rather extreme geometries of
dimensions comparable with domain wall thickness, such as
ultrathin films, short-period superlattices and ferroelectric
nanodots.\cite{Fu03,Scott05,Prosandeev06,Rodriguez09} 
Domain walls in ferroelectric films with an order of magnitude larger
thickness (of the order of 100\,nm) are expected to bear mostly
bulk material properties. Nevertheless, the necessity of stress
and depolarization charge compensation at 100\,nm scales implies
presence of a much higher density of domain walls and this can also
make conditions for certain nanoscale-specific domain
arrangements, different from typical bulk domain structures.

The key factor dictating the domain structure type in epitaxial
ferroelectric films is the lattice misfit strain. In general,
compressive stress favors so-called c-domain states with
out-of-plane polarization orientation, while tensile stress favors
a-domain states with in-plane polarization orientation. However,
the relative weights of ferroelastic domain variants in the domain
structure is not the only mechanism for relieving the misfit
stresses. Typically, in case of epitaxial ferroelectric films with
thicknesses of 50-500\,nm, a considerable amount of epitaxial
stress is relaxed by formation of dislocations. Therefore, the
final domain structure may depend on a number of other factors,
such as growth temperature and annealing
history.\cite{Book2,Speck94I,Speck94II,Alpay98III}
Perspective of practical applications of such thin ferroelectric
films, for example in so-called MEMS devices, obviously calls for
a better insight in the processes governing the formation of their
domain texture. As a contribution to this problem, we present here
investigations of PbTiO$_3$ (PTO) epitaxial thin films grown on two
different high quality  rare earth scandate 
crystal substrates. 
Interestingly, although these films were prepared simultaneously
in the same MOCVD deposition process, the resulting nanodomain
architecture of these thin films is drastically different.

\section{Experimental}
The PTO thin film samples for this investigation were grown by
pulsed injection metal-organic chemical vapor deposition (MOCVD) technique 
on commercial high-quality
(110)-oriented TbScO$_3$ (TSO) and SmScO$_3$ (SSO) single crystal
substrates developed for epitaxial deposition of perovskite
materials.\cite{SchlomAnnu07} Deposition took place at about
650\,$^{\circ}$C (for other details, see e.g.
Refs.\,\onlinecite{barta1, barta2}). Films were
deposited  on both substrates  simultaneously to avoid possible differences in the
deposition history. Preliminary room-temperature X-ray diffraction
taken just after the growth clearly indicated single-crystalline
epitaxy with preferential $c$-domain orientation in the film grown
on TSO (less than 10 volume percent of residual $a$-domains) and
preferential $a$-domain orientation in the film grown on SSO (with
about 6 volume percent of residual $c$-domains). Thickness of the
resulting PTO films (about 320\,nm) was determined on another film
(grown on Si substrate during the same deposition), which was etched 
with concentrated HF to form a step for thickness measurement by profilometry. 
All present measurements were done at ambient
conditions on as-grown films, no thermal cycles were done after.

The piezoresponse force microscopy (PFM) measurements were 
conducted using the atomic force microscope (AFM)
of the Ntegra spectra apparatus operated with a conductive
tip (TiN-coated n-doped silicon cantilever) in a contact
mode. Using silver paste, the samples were carefully glued to a special plate provided with a spring contact. 
The frequency of the alternating voltage $V_{ac}$ was set 
to a value of $f\approx$ 15\,kHz and the amplitude to $V=$ 5\,V. The mechanical
response of the cantilever-tip-surface system detected in a
standard way was amplified and analyzed with an external SR830 DSP Lock-In Amplifier.

Polarized Raman spectra were measured using Renishaw Raman RM-1000
 Micro-Raman spectrometer with a CCD detection.
 The experiments were performed in backscattering geometry in 
the 20-900\,cm$^{-1}$ range. The 514.5 nm 
line of an Ar$^+$ ion laser was focused to a spot size of about 2\,$\mu$m.
 The VV and HV polarization configurations correspond to
situations where the input and output light polarizations are parallel and crossed, respectively.
TbScO$_3$ and SmScO$_3$ substrates have several Raman modes within the 
region of interest. Consequently, the spectra contained modes 
of the substrate as well as of the material and were difficult to analyze. To obtain pure film 
spectra, substrate spectra were measured separately  and the substrate contribution was subtracted from film+substrate spectra.

\section{Results}
\begin{figure}[ht]
\includegraphics[width=50mm]{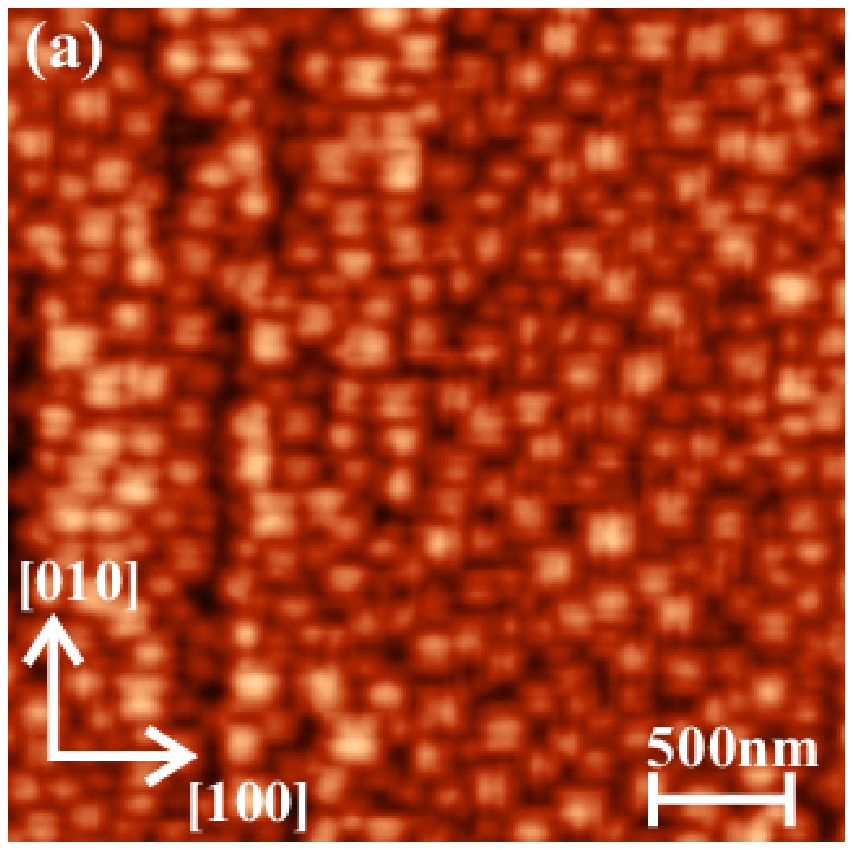} 

\includegraphics[width=50mm]{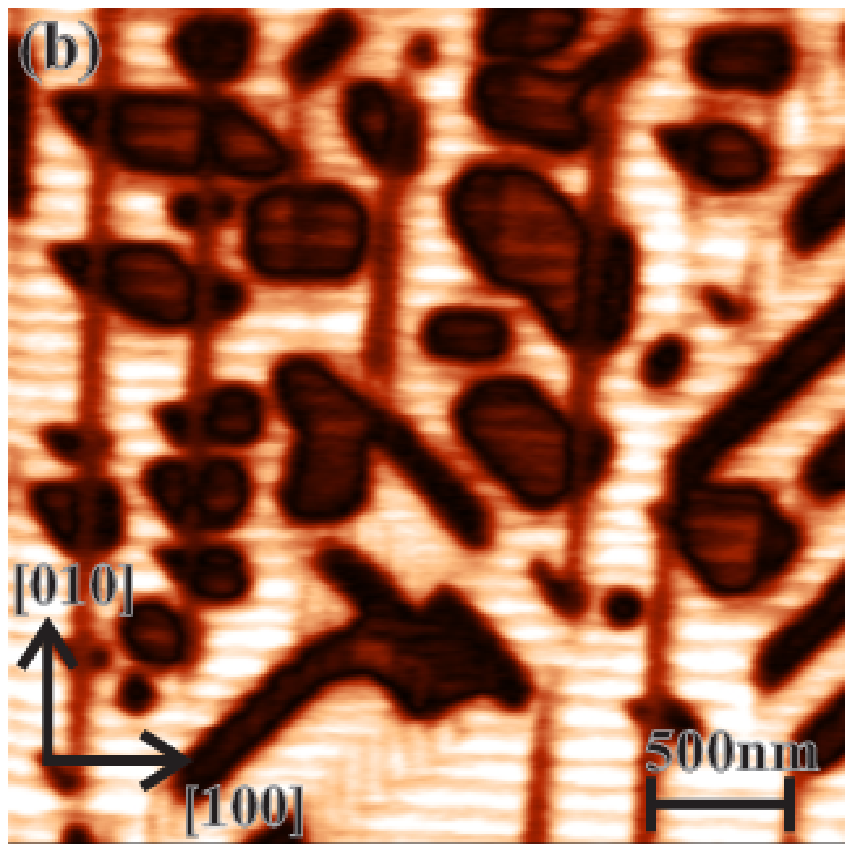}

\includegraphics[width=50mm]{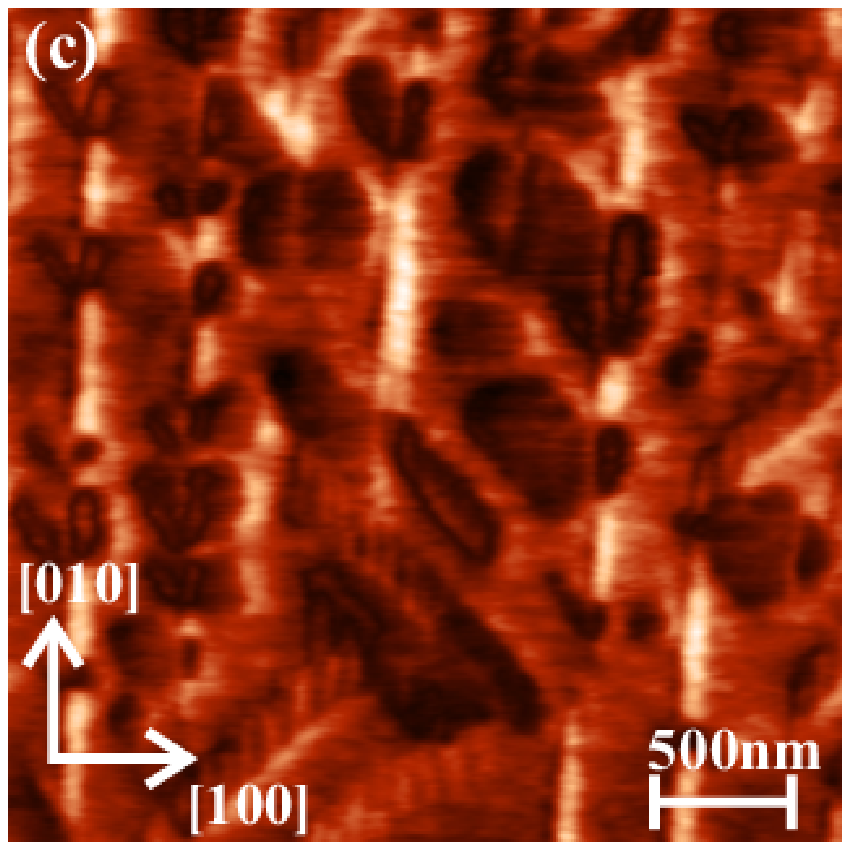}
 \caption{(Sorry, this preprint contains  lower-resolution images only)
(a) Topography,  (b) vertical and (c) lateral PFM images of 320\,nm thin PbTiO$_3$
film grown on TbScO$_3$ substrate.
 Edges of the scanned area are roughly parallel to pseudocubic axes of the substrate and the epitaxial film as well.} \label{figTSO}%
\end{figure}

\begin{figure}[ht]
\includegraphics[width=50mm]{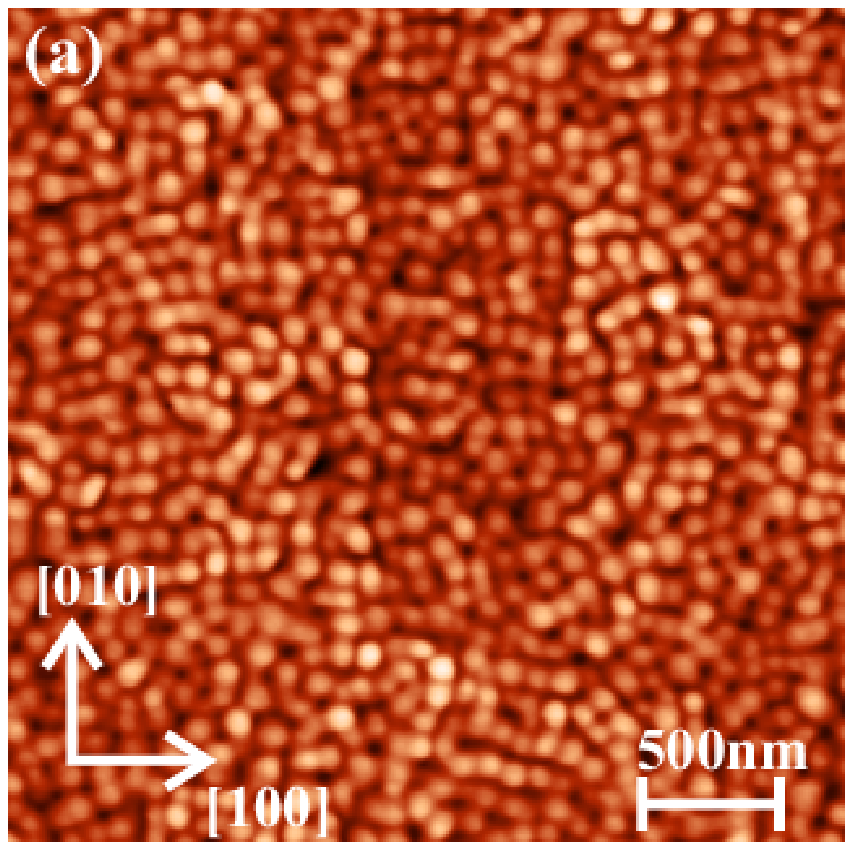}

\includegraphics[width=50mm]{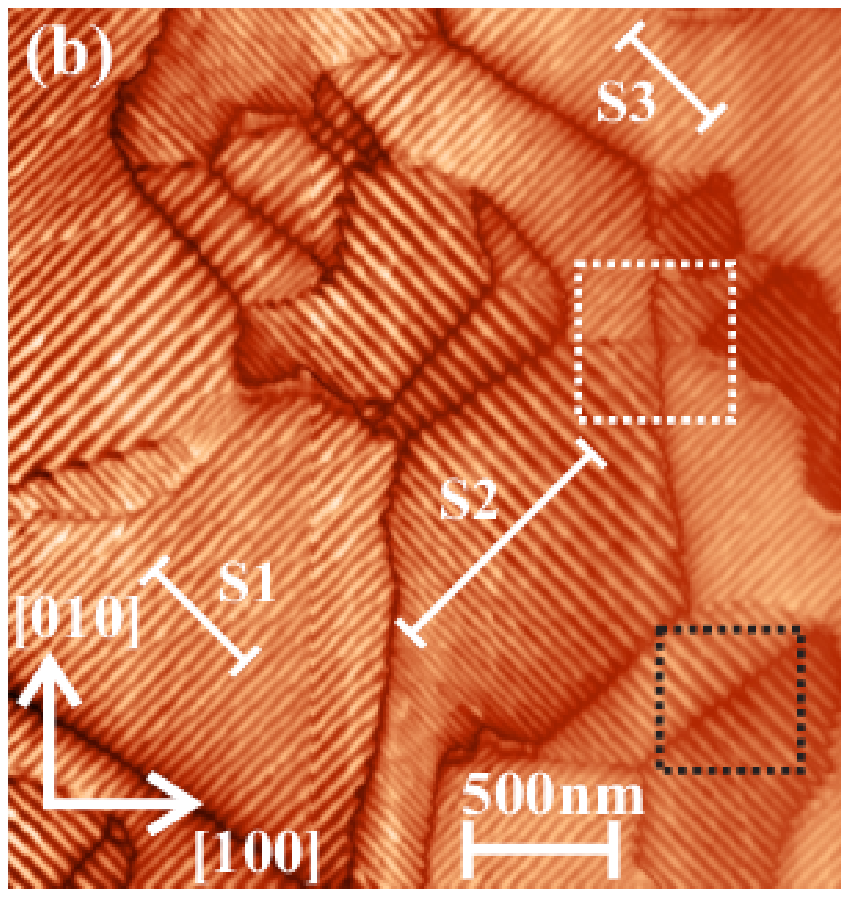}

\includegraphics[width=50mm]{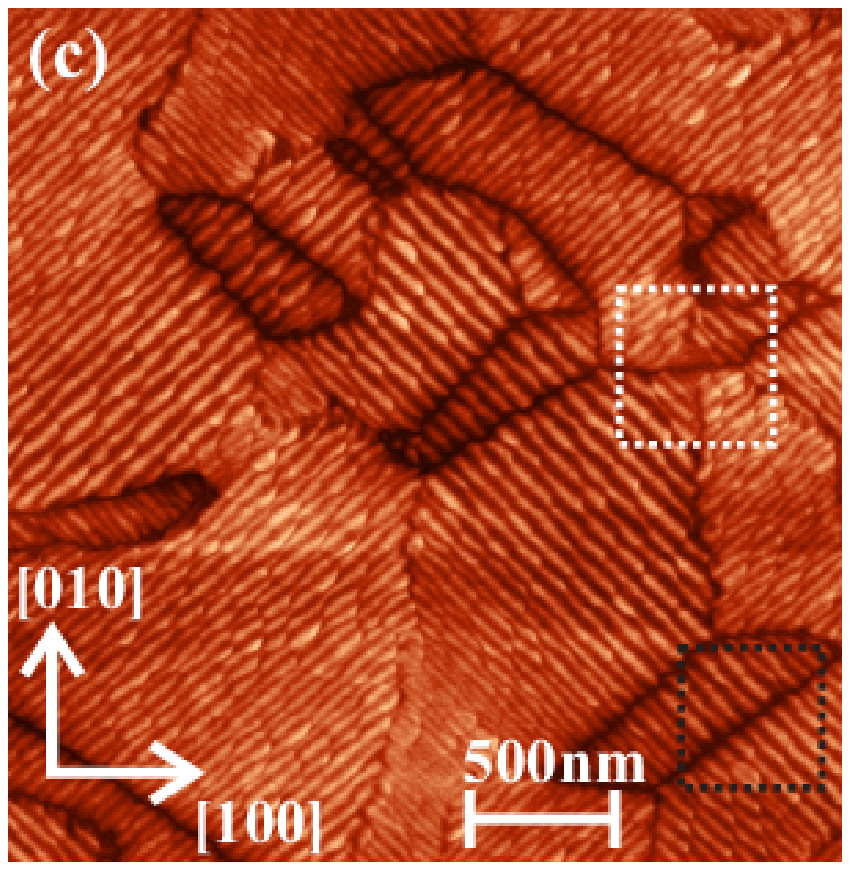}
 \caption{(Sorry, this preprint contains  lower-resolution images only)
(a) Topography, (b) vertical and (c) lateral PFM images of 320\,nm thin PbTiO$_3$ film on SmScO$_3$ substrate.
 Edges of the scanned area are roughly parallel to pseudocubic axes of the substrate and the epitaxial film as well. The black frame indicates the area 
shown enlarged in Fig.\,\ref{figSSO_DW45} and the white frame indicates the area shown enlarged in Fig.\,\ref{figSSOvortex}. 
White lines S1, S2 and S3 refer to cross-sections of the PFM image shown in Fig.\,\ref{figSSOperiod}.} \label{figSSO}%
\end{figure}

Typical AFM images of the investigated PTO films are shown in
Fig.\,\ref{figTSO} and Fig.\,\ref{figSSO}. The topographic imaging
mode has shown a gentle surface corrugation. The morphology of
this roughness can be described as a dense disordered pattern of
few nm high circular protuberances, indicating an island growth
mechanism, frequently encountered in MOCVD
deposition.\cite{barta4} The surface topography of the film grown
on TSO substrate is very similar to the film grown on SSO
substrate. However, the PFM images are quite different. The PFM
images of PTO/TSO film reveal dominantly $c$-domain signal with
narrow minor $a$-type domains in a $c/a/c/a$ arrangements,
 while the film grown on SSO substrate is mostly showing only $a$-domains
  arranged in $a/a/a/a$ stripes revealing fine structure of (110) and (1$\bar{1}0$) oriented 90-degree walls.

For example, the basic motif of the vertical-signal PFM image of
PTO film on TSO shown in Fig.\,\ref{figTSO} appear to be the
 "Swedish ladder" pattern formed by quasi-regular alternations of about 100\,nm wide
$a$-c-domain stripes with an order of magnitude narrower
$c$-domain stripes (faint-contrast "bars" of the Swedish ladder).
This pattern is very similar to the domain arrangements reported for
PTO films grown on Nd-doped SrTiO$_3$ (STO)
substrates.\cite{LeeChoiLeeBaik01} Another,
less dense set of about 150\,nm thick stripes (vertical on
Fig.\,\ref{figTSO}, intermediate darkness) has been also observed
in PTO films grown on STO
substrates\cite{LeeChoiLeeBaik01} and it
can be assigned to a minor fraction of $a/a/a/a$
lamellae.\cite{Nakaki08,Nakaki09,Utsugi09}  Finally, the
kidney-shaped (dark) islands with about 100\,nm diameter
correspond to the inverted structure with opposite overall
spontaneous polarization. Consequently, the borders of these
islands are mostly formed by 180-degree domain walls. These
180-degree domain interfaces help to minimize the depolarization
fields normal to the film. This is likely the reason why they 
do not form spontaneously on conductive substrates like Nd-doped
STO. However, similar density and pattern of 180-degree walls was
observed in PTO films grown on LaAlO$_3$ (LAO) substrates.\cite{Simon11}

The orientation of polarization in PTO films grown on SSO can be
identified by the analysis of the PFM signal in the vertical and
the lateral mode. The long axis of the AFM cantilever was parallel
to the [010] direction indicated in Fig.\,\ref{figSSO}, so that
the contrast of the lateral PFM images is mostly given by the
[100] polarization component. Since the polarization is almost
exclusively in-plane oriented in most of the images, the vertical
PFM images give mainly contrast along the [010] axis (due to the
cantilever buckling effect).\cite{Buckling1,Buckling2,Buckling3}
Comparison of lateral and vertical mode images taken in the same
area clearly confirms presence of about 0.5-1 micron size "coarse
domains" formed by regular nanodomain $a/a/a/a$ twinned areas with
about 20\,nm wide $a$-domain stripes with polarization oriented
alternatively along [100] and [010] axes. These patterns are very
similar to the domain structure seen on PTO films grown on
KTaO$_3$ (KTO) crystal substrates.\cite{LeeChoiLeeBaik01} It is natural to expect that these
stripes are separated by mechanically and electrically compatible
head-to-tail 90-degree domain walls normal to either [110] or
[1$\bar{1}0$].


\begin{figure}[ht]
\includegraphics[width=75mm]{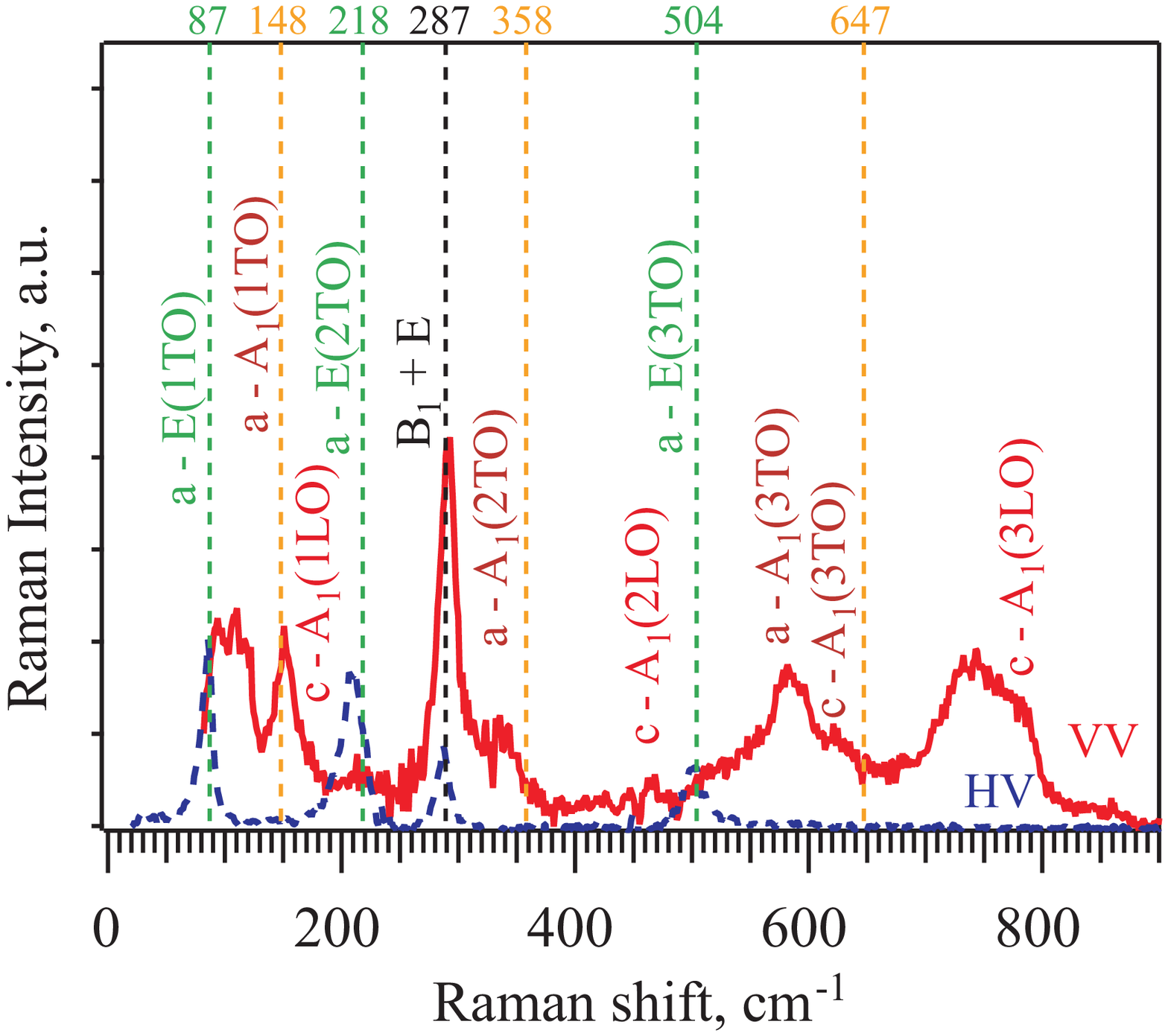}
 \caption{(Sorry, this preprint contains  lower-resolution images only)
Polarized Raman spectra of 320\,nm thin PbTiO$_3$ film on TbScO$_3$ collected in crossed (HV) and parallel
(VV) polarization configurations. Vertical dashed lines correspond to PTO single crystal modes. } \label{figTSORaman}%
\end{figure}

\begin{figure}[ht]
\includegraphics[width=75mm]{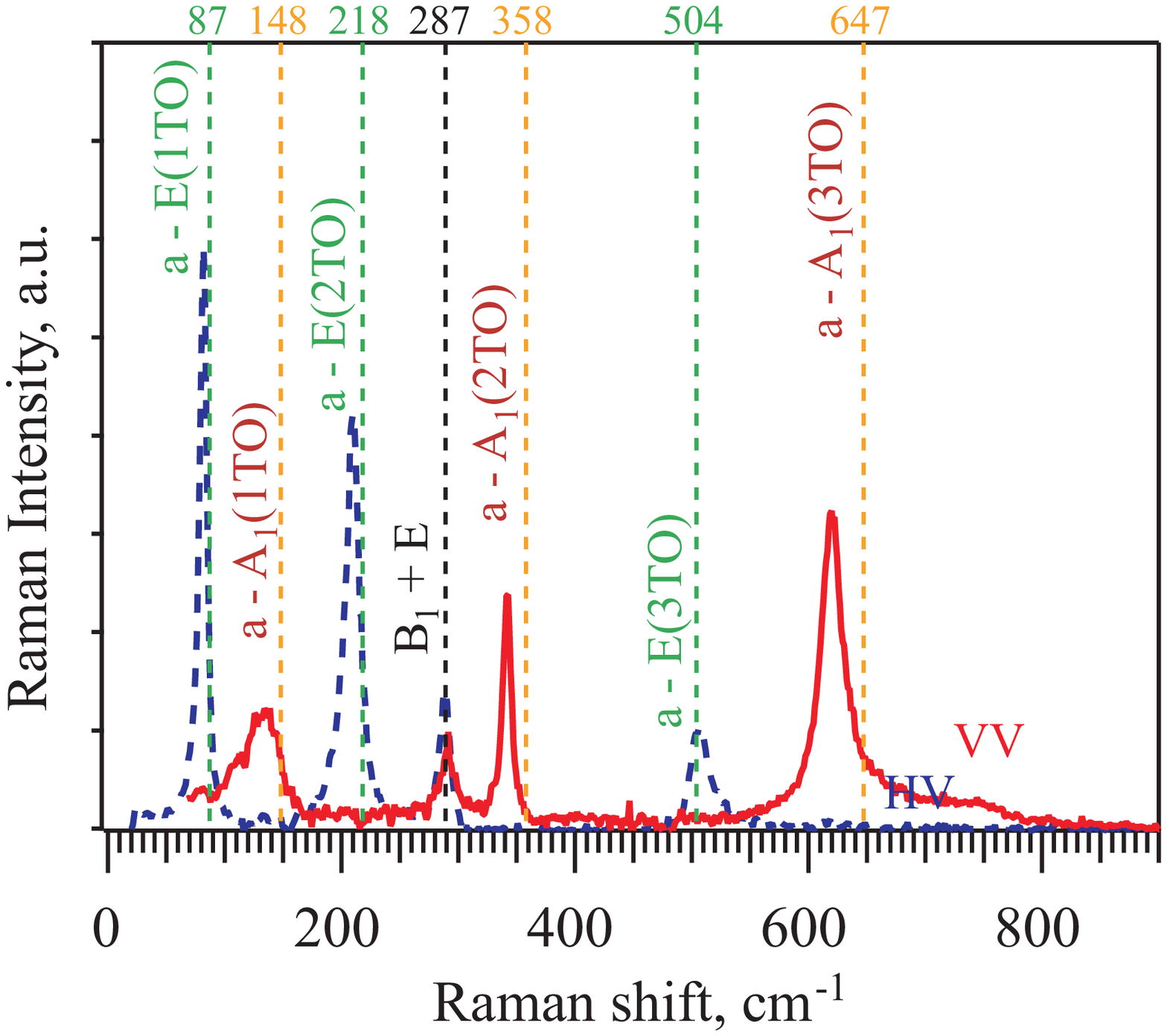}
 \caption{(Sorry, this preprint contains  lower-resolution images only)
 Polarized Raman spectra of 320\,nm thin PbTiO$_3$ film on SmScO$_3$ recorded in crossed (HV) and parallel (VV) polarization configurations.
 Vertical dashed lines correspond to PTO single crystal modes. } \label{figSSORaman}%
\end{figure}

 The preferential $c$-domain and $a$-domain occurrence in PTO films
grown on TSO and SSO films, respectively, can be also documented
by polarized Raman spectroscopy.
 Typical backscattering Raman spectra taken from the PTO/TSO sample surface in
$z(xx)\bar{z}$ and $z(xy)\bar{z}$ geometry (after subtraction of the substrate signal) are shown in Fig.\,\ref{figTSORaman}. 
Similar spectra for  PTO/SSO sample are shown in Fig.\,\ref{figSSORaman}.
 Modes can be assigned by comparison with previous measurements on
PbTiO$_3$ films.\cite{barta3} The basic rule here is that according to the
standard selection rules for symmetric Raman tensors, the pure
$A_1$ phonon modes (with dynamical charge fluctuating along the
tetragonal axis) should be active in the $z(xx)\bar{z}$ spectra
but not in the $z(xy)\bar{z}$ spectra, while the pure $E$ modes
(with dynamical charge fluctuating perpendicular to the tetragonal
axis) should be active only in the $z(xy)\bar{z}$ spectra. The
presence of strong Raman bands close to $A_1(LO)$ frequencies of
bulk PbTiO$_3$ (for example, the $A_1(3LO)$ band near 770 cm$^{-1}$)
thus suggests a large volume of areas with tetragonal axis normal
to the film ($c$-domains). In contrast, the $A_1(3LO)$ band is
barely seen in the similar spectrum taken from the PTO film grown
on SSO substrate, confirming in this way a negligible volume of
the $c$-domain in such sample. Moreover, $E$ modes from
$c$-domains should not be active either in the $z(xx)\bar{z}$
or in the $z(xy)\bar{z}$ spectrum. Therefore, only the $E$
modes from $a$-domains are observed in the adopted backscattering
geometry. Since there is only a minute fraction of $a$-domains
in TSO-grown film, the $z(xy)\bar{z}$ spectrum of the TSO-grown
sample is indeed quite weak.

\section{Discussion}

Using three different techniques (X-ray diffraction, PFM
imaging and Raman spectroscopy) we have seen that the tetragonal axis of the
PTO has a dominant in-plane orientation in case of films grown on
 SSO substrate, while, on the other hand, it is clearly preferentially perpendicular to the film
 in case of the TSO substrates. The room-temperature pseudo-cubic lattice
 parameters of both substrates  (about 3.960 and 3.985\,{\AA}  for TSO and SSO,
 resp.)\cite{Uecker08} are both between the  $a$ and $c$ room-temperature lattice
 parameters of bulk tetragonal PTO (about 3.90 and
 4.155\,{\AA}).\cite{Mabud79} Therefore, a perfect atomic epitaxy should lead to
 a tensile epitaxial stress in $c$-domains and biaxial in-plane
 strain in $a$-domains (compressive local stress in PTO lattice along its in-plane-oriented tetragonal axis,
 tensile stress in the other direction in the plane of the film).

 However, these naive considerations neglect the effect of other
 factors like thermal expansion and formation of dislocations during the deposition.
 Therefore, it is instructive to examine possible stress-induced phonon
 frequency shifts in the recorded Raman spectra. For this purpose,
 we have marked the positions of the known stress-free bulk phonon
 frequencies directly in  our Raman spectra. Phonon frequency shifts are particularly clear
 in the case of the film grown on SSO: all $A_1$(TO) modes in the dominant $a$-domain configuration show a considerable frequency
 downshift with respect to the bulk mode frequencies (about 30\,cm$^{-1}$).
 Such shifts are quite unusual for a 320\,nm thick film; in fact,
comparable frequency shifts were
 reported in the case of $c$-domains in PTO films grown on LaAlO$_3$, but only when the
 films were about one order of magnitude thinner.\cite{barta4}
 As it is known  that uniaxial compression of PTO along the tetragonal axis decreases
 all $A_1$(TO) modes (see, for example
 Ref.\,\onlinecite{Marton13}), the observed Raman shifts
 suggest that $a$-domains are really under compression  along the local tetragonal axis there.
 On the contrary, $E$(TO) modes in the dominant $a$-domain configuration show much smaller shifts,
 what could be an indication of the expected local biaxial
 in-plane epitaxial strain.

 Although we do not fully understand the decisive criteria for
 the formation of such a nicely strained PTO film with $a/a/a/a$ structure, we believe that the essential difference
 between the TSO and SSO substrate in this respect is that the
 high-temperature PTO film growth on SSO substrate occurs in conditions of a slight tensile straining. 
 In fact, the only other case of PTO epitaxial film with a very similar
 $a/a/a/a$ domain structure known to us is that of PTO film grown on KTO substrate,\cite{LeeChoiLeeBaik01}
 which happens to have a lattice constant very similar to SSO (room-temperature lattice parameter of KTO is about 3.99\,{\AA}).\cite{SchlomAnnu07} 
On the contrary, the  TSO substrate and most of the other popular substrates (STO, LAO,
(La$_x$Sr$_{1-x}$)(Al$_y$Ta$_{1-y}$)O$_3$) have a smaller lattice constant than the cubic PTO at
 650\,$^{\circ}$C.

 Indeed, pseudocubic lattice parameter of bulk SSO at 650\,$^{\circ}$C $a'_{SSO}$= 4.01\,{\AA} 
(evaluated using linear thermal expansion coefficient and room
 temperature lattice constant from Ref.\,\onlinecite{Uecker08}), while that of bulk TSO at 650\,$^{\circ}$C
 is only $a'_{TSO}$= 3.96\,{\AA} (also evaluated using linear thermal expansion
 coefficient and room temperature lattice constant from Ref.\,\onlinecite{Uecker08}).
 Therefore, the lattice parameter of bulk cubic PTO at 650\,$^{\circ}$C  $a'_{PTO}$= 3.99\,{\AA} 
(extrapolated from data of Ref.\,\onlinecite{Mabud79}) falls in between that of TSO and
 SSO ($a'_{TSO}<a'_{PTO}<a'_{SSO}$) and perfect epitaxial matching would imply that
 thin PTO films grown on TSO and SSO exhibit at growing conditions
 compressive and tensile biaxial in-plane strain respectively. Such strain
 is rather large in case of the 300 nm thick film, and, most likely, it is
 partly reduced by formation of dislocations.

At and below the ferroelectric phase transition (near 500\,$^{\circ}$C), the
 "natural" lattice parameters of bulk PTO are modified by spontaneous
 strain, and this has most significant impact on the local lattice
 mismatch variation on cooling. Since the spontaneous lattice strain is
 considerable in bulk PTO ($c$ parameter is by 6 percent larger than the $a$
 parameter at room temperature), the average in-plane lattice parameter of
 PTO film obviously strongly depends on the volume ratio of domains with
 different $c$-axis orientation. At a first approximation, the $a/a/a/a$
 twinned area would have average in-plane lattice parameter $(a+c)/2$, i.e.
 about 4.03\,{\AA} at room temperature, which is by 3 percent more than the
 natural in-plane lattice constant of a single domain PTO film with $c$-axis vertical 
 ($a$=3.90\,{\AA} at room temperature). The actual domain formation process probably 
involves nucleation and motion of ferroelectric domain walls as well as lattice dislocations, but its
details are not obvious to us.

\begin{figure}[ht]
\includegraphics[width=40mm]{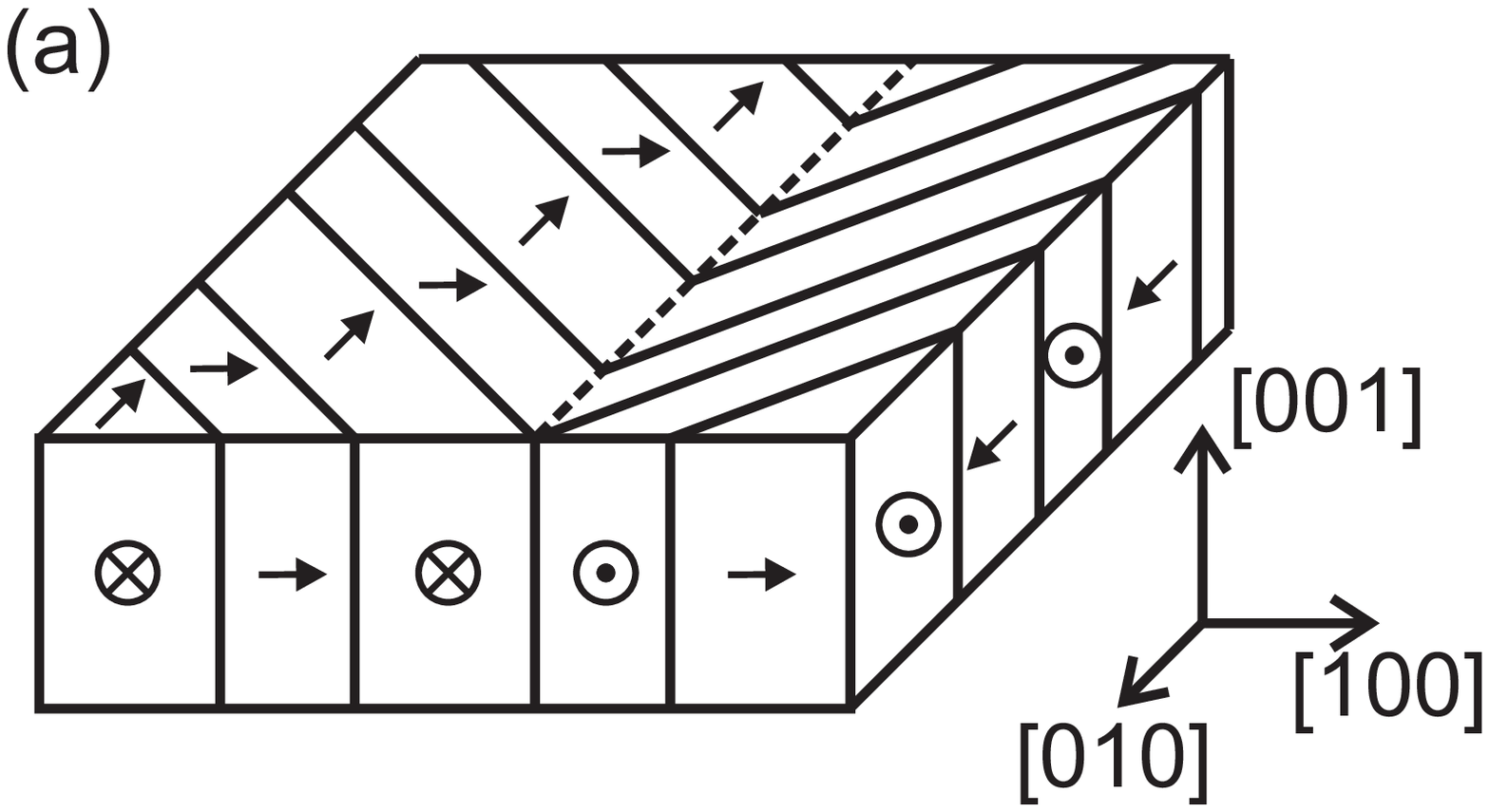} \includegraphics[width=40mm]{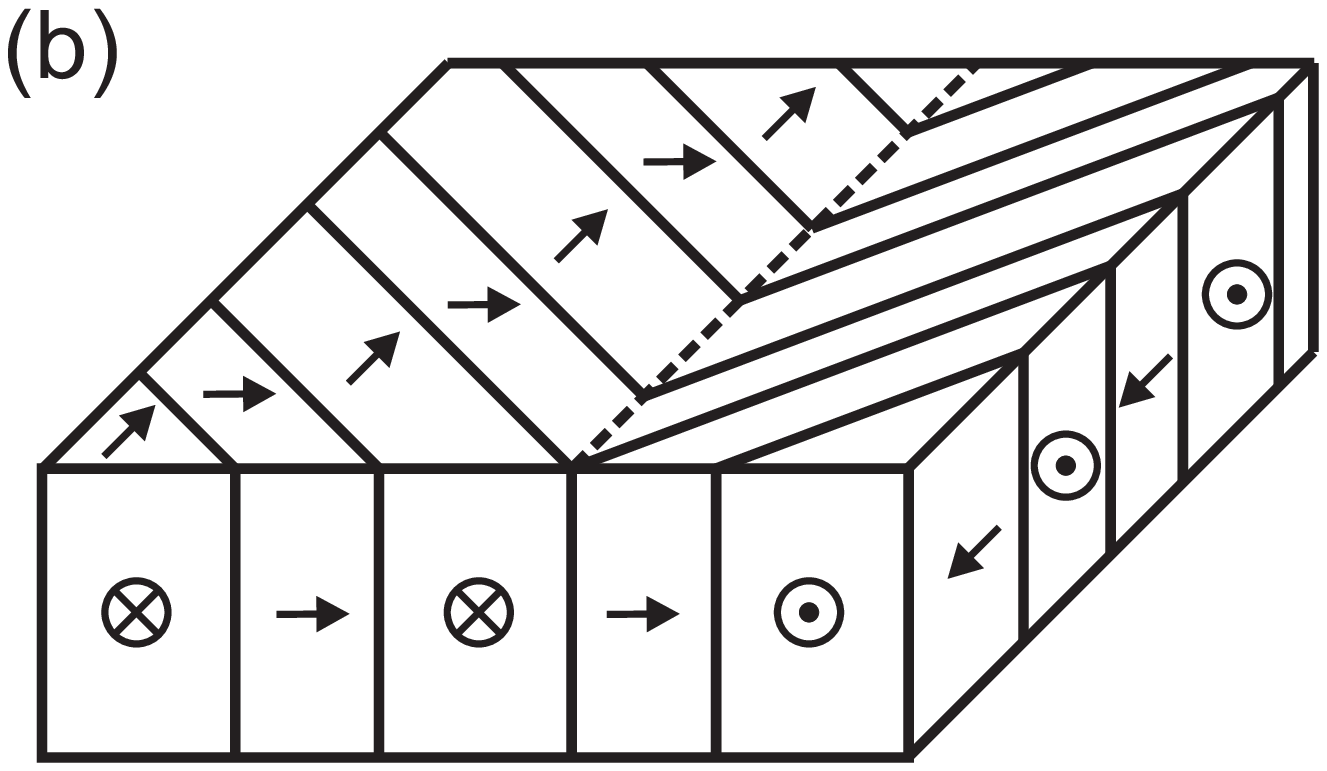}
 \caption{(Sorry, this preprint contains  lower-resolution images only) Schematic suggestions of domain structure arrangement at the [010]-type mesoscopic 
domain boundary between two differently twinned area of
 $a/a/a/a$ domain structure. The right-hand-side mesoscopic domain boundary
 is formed by 90$^\circ$-domain walls with unusual crystallographic orientation (forbidden in bulk PbTiO$_3$).
Note that [010] as well as [110]-type mesoscopic domain boundaries are quite frequent in the 320\,nm
 thin PbTiO$_3$ film on SmScO$_3$ (see Fig.\,\ref{figSSO}(b)).} \label{figSSOdomain}%
\end{figure}

 It is also interesting to note that the $a/a/a/a$ domain
 structure observed in the PTO films grown on SSO happens to be
 similar to the domain structures observed in the focused-ion-beam-cut
 free-standing lamellae of BaTiO$_3$ single crystals.\cite{Schilling06} In both
 cases, the present PTO film and BTO lamellae, the areas of quasi-regular $a/a/a/a$ pattern motifs are
 forming larger-scale "mesoscopic" domains, separated by narrow
 interfaces, which can be considered as "mesoscopic domain
 walls". Like the true ferroelectric boundaries, these mesoscopic domain
 walls shows a clear directional preferences - they tend to be
 parallel to the [100] and [010] directions or close to [110] and
 [1$\bar1$0] directions. Idealized microstructure of such [010]
 mesoscopic boundary between two $a/a/a/a$ domains is sketched in
 Fig.\,\ref{figSSOdomain}. The mesoscopic boundary in
 Fig.\,\ref{figSSOdomain}(a)  in fact corresponds to a sequence of
(charge and mechanically compatible) 180-degree ferroelectric
domain walls with [010] orientation
 and "no-wall" regions, where local domain state is not changed at
 all. On the other hand, the mesoscopic boundary in
 Fig.\,\ref{figSSOdomain}(b) is formed by a sequence of 90-degree
 walls, which are locally in a electrically and mechanically incompatible
 arrangement,\cite{Janovec69, Marton10} even though the overall mesoscopic
 boundary can be in both cases considered as a charge-neutral head-to-tail boundary.
 We have not seen any clear
 preference for the structure of Fig.\,\ref{figSSOdomain}(a) in our
 images, probably also because it requires identical twinning
 period in the adjacent mesoscopic domains. Nevertheless, the TEM
 image of the corresponding mesoscopic boundary in PTO film grown on
 KTO seems to support  preference for the structure of Fig.\,\ref{figSSOdomain}(b) (see Fig.\,10 of Ref.\,\onlinecite{LeeChoiLeeBaik01}).

\begin{figure}[ht]
\includegraphics[width=50mm]{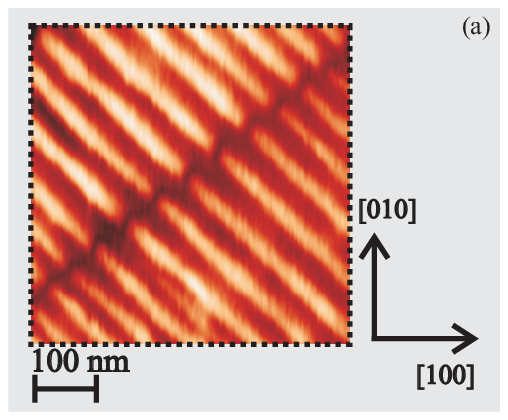}

\includegraphics[width=50mm]{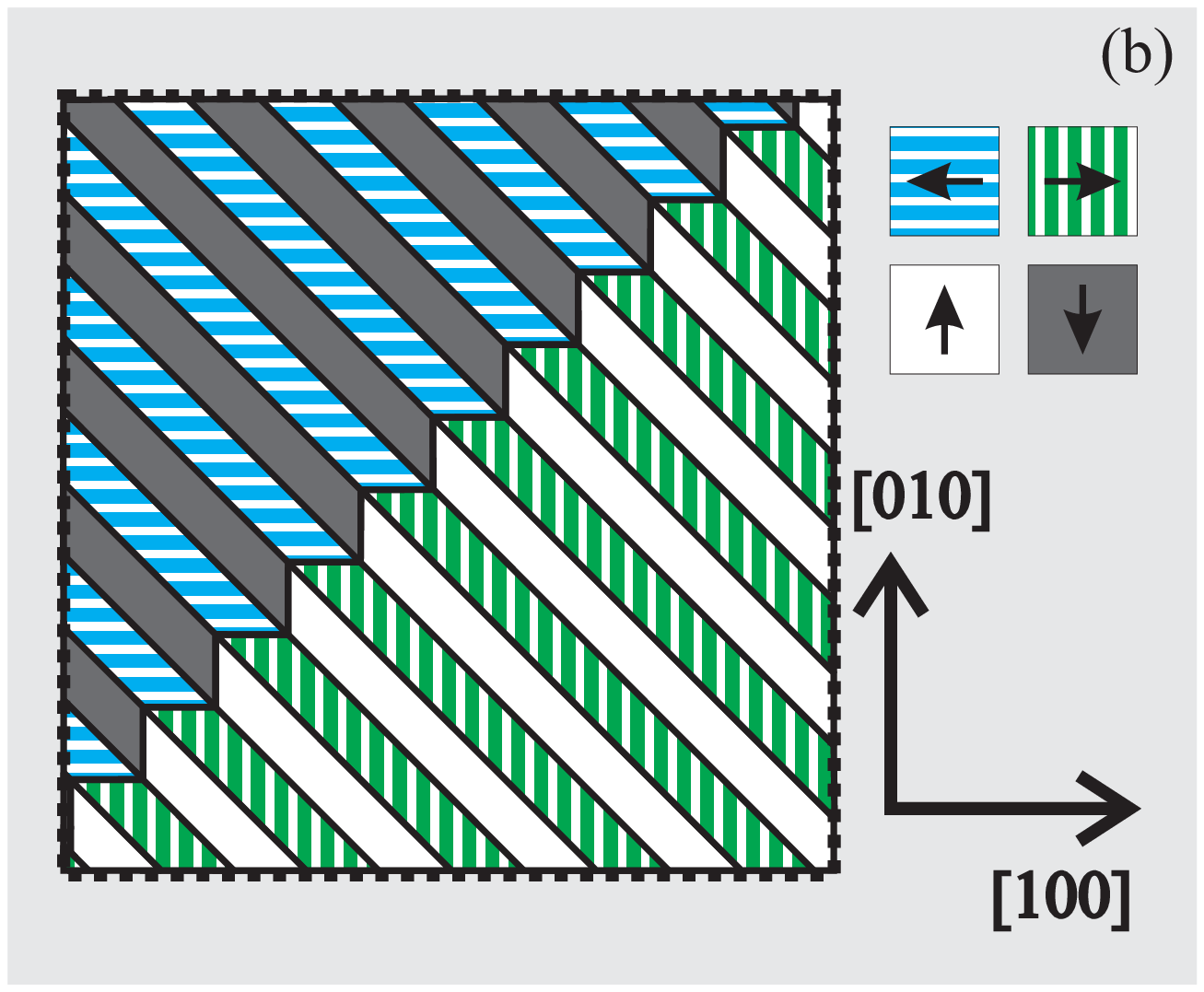}\\

 \caption{(Sorry, this preprint contains  lower-resolution images only)
 (a) Enlarged portion of Fig.\,\ref{figSSO}(b), showing a mesoscopic domain boundary oriented along [110] direction. 
 (b) Schematic suggestion of domain arrangements with electrically and mechanically compatible walls.} \label{figSSO_DW45}%
\end{figure}

 The structure of mesoscopic  boundaries oriented along [110] direction is shown in Fig.\,\ref{figSSO_DW45},
 obtained by zooming from Fig.\,\ref{figSSO}. Here the boundary is clearly formed by an array of wedge-type
 nanodomain terminations, rather than by any type of known flat compatible domain walls.
 This observation strongly reminds the "bundle
 boundary" observed in BaTiO$_3$ lamellae.\cite{McGillyNano10} In principle, however, the mesoscopic domain 
 boundary could be formed by mechanically compatible uncharged 180-degree domain walls, as proposed in Fig.\,\ref{figSSO_DW45}(b).

\begin{figure}[ht]
\includegraphics[width=50mm]{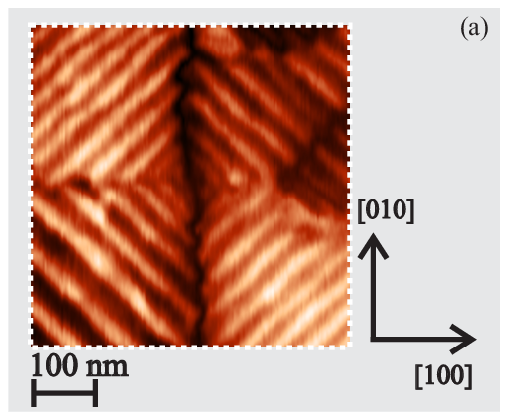}

\includegraphics[width=50mm]{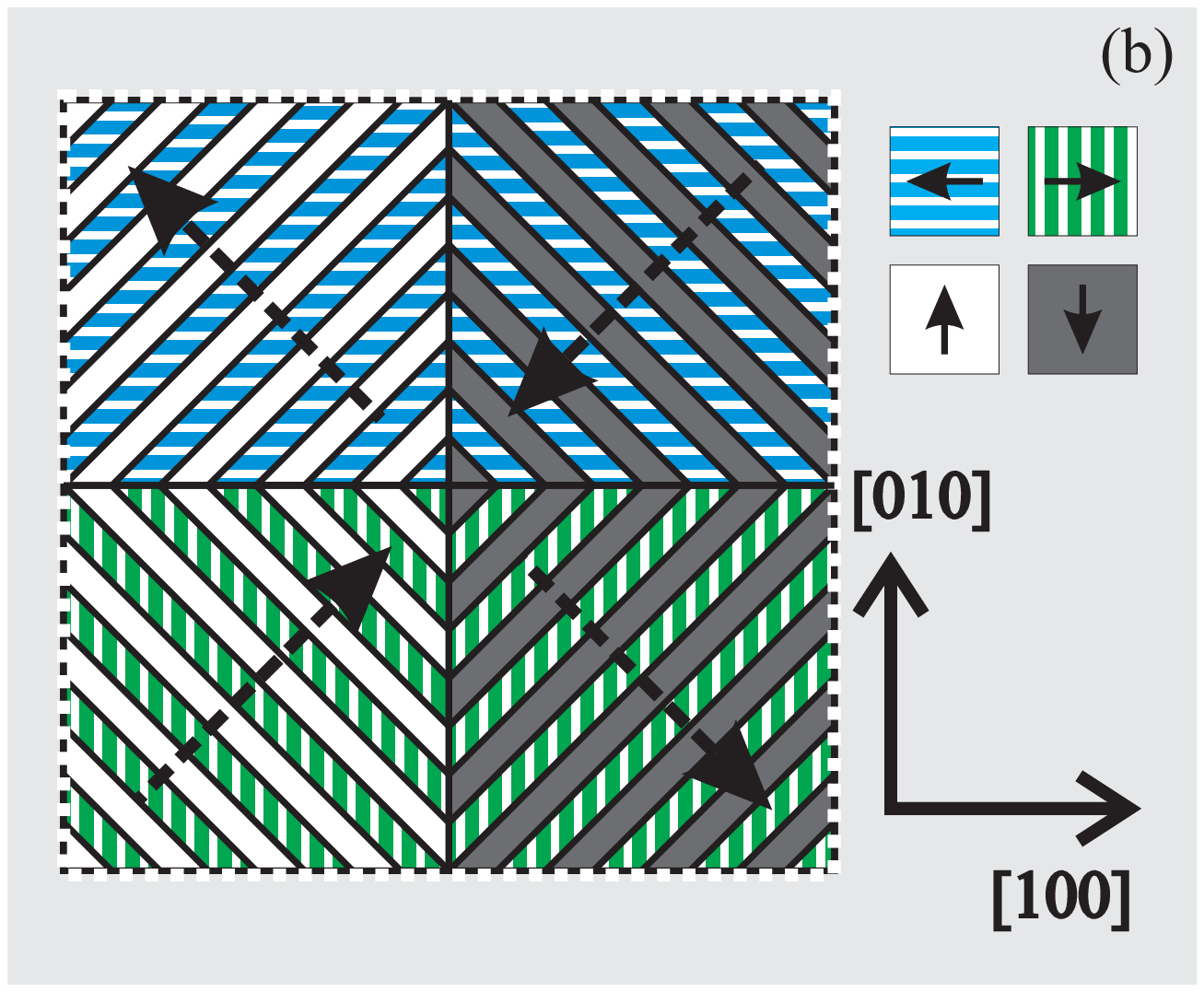}\\

 \caption{(Sorry, this preprint contains  lower-resolution images only)
Quadrant-quadruple domain pattern.  
 (a) Enlarged portion of Fig.\,\ref{figSSO}(b), showing the intersection of two mesoscopic domain walls oriented along [100] and [010] direction.
 (b) Schematic suggestion of domain structure assignment. 
 Black dashed arrows in (b) correspond to macroscopic polarization directions.} \label{figSSOvortex}%
\end{figure}

 In the present experiment we also observed several mesosocopic domain intersections.
 The intersection of two mesoscopic domain walls oriented along [100] and [010] direction
 forms an interesting object which is probably quite analogous to the "quadrant-quadrupole"
 mesosocopic domain structures known from BaTiO$_3$ lamellae.\cite{SchillingPRB11} 
 Example of such quadrant-quadruple mesoscopic domain structure, obtained by zooming from
 Fig.\,\ref{figSSO}(b), is shown in Fig.\,\ref{figSSOvortex}.

\begin{figure}[ht]
\includegraphics[width=65mm]{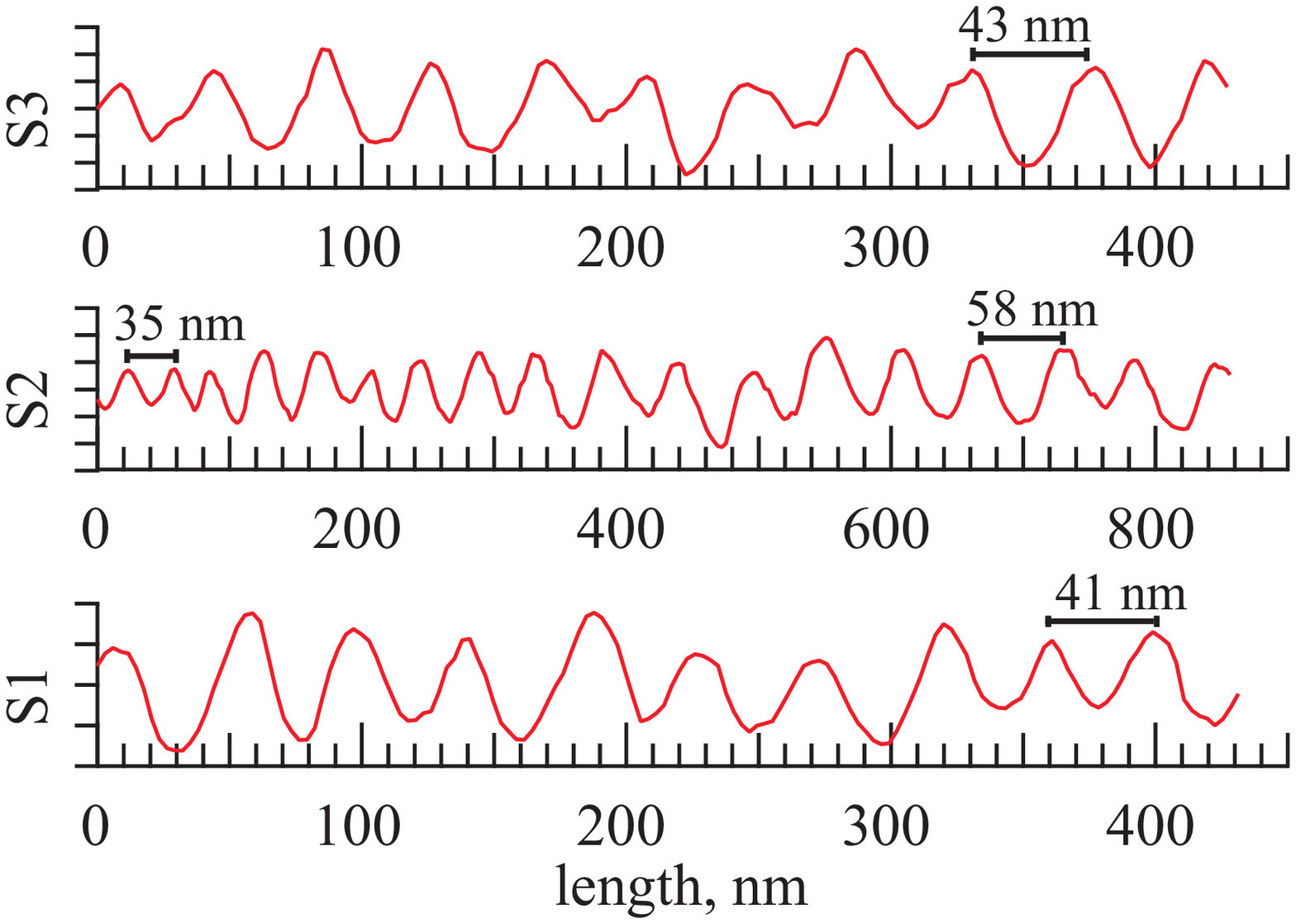}
 \caption{(Sorry, this preprint contains  lower-resolution images only)
 Cross-sections of the vertical PFM image along the lines marked  in Fig.\,\ref{figSSO} by white lines S1, S2 and S3.} \label{figSSOperiod}%
\end{figure}

Finally, Fig.\,\ref{figSSOperiod} shows several representative
scans through PFM data, allowing to extract quantitative
information about nanodomain periods. The period fluctuates
between mesoscopic domains as well as within single domains; but
the average nanodomain size of 20-30\,nm roughly agrees with the
values expected from Kittel's law for a 320\,nm thick
plate.\cite{Catalan07,streiff02}

\section{Conclusion}

Epitaxial films of ferroelectric PbTiO$_3$, about 320\,nm thick, 
have been simultaneously deposited by MOCVD technique on two
different scandate single crystal substrates. Subsequent X-ray,
PFM and Raman confirmed that PTO films grown on TSO have
preferentially $c$-domain orientation, while films grown on SSO
have dominantly $a$-domain orientation (in-plane orientation of
spontaneous polarization). The striking difference between the two
domain structures is assigned to the opposite sign of the
epitaxial misfit strain at the deposition temperature. The
strained PTO films grown on SSO substrate shows interesting domain
arrangements analogous to the structures reported in free-standing
BaTiO$_3$ single crystal lamellae. We believe that these results
will be useful for understanding and design of epitaxial
ferroelectric films with thickness of about 300\,nm.

\begin{acknowledgments}
This work was supported by the Czech Science Foundation (Projects
Project GACR P204/10/0616
 and 202/09/H0041) and EGIDE (Grant No. GILIBERT 25536 WF). 
In addition, the contribution of Ph.D. student F. Borodavka has been
supported by Czech Ministry of Education (project
SVV-2012-265303).
\end{acknowledgments}

\end{document}